\documentclass[pre,preprint]{revtex4-1}

\usepackage{amsmath}  
\usepackage{amsfonts} 
\usepackage{graphicx} 
\usepackage{subfig}
\usepackage{enumitem} 
\usepackage{xcolor} 
\usepackage[utf8]{inputenc} 
\usepackage[normalem]{ulem} 

\begin{document}

\title{Prime stars multiplexes}

\author{$^{1,2}$Alfonso Allen-Perkins}
\email{alfonso.allen.perkins@gmail.com}
\author{$^{1,3}$Roberto F. S. Andrade}
\email{randrade@ufba.br}
\affiliation{$^{1}$Instituto de F\'isica, Universidade Federal da Bahia, 40170-115 Salvador, Brazil.}
\affiliation{$^{2}$Complex System Group, Universidad Polit\'ecnica de Madrid, 28040-Madrid, Spain.}
\affiliation{$^{3}$Centre for Data and Knowledge Integration for Health (CIDACS), Instituto Gon\c{c}alo Muniz, Funda\c{c}\~{a}o Oswaldo Cruz (FIOCRUZ), 41745-715 Salvador, Brazil}

\date{\today}

\begin{abstract}
This work investigates the class of prime star multiplexes, in which each of its  layers $i$, $i=1,2, \ldots M$, consists of a regular cycle graph where any node has $2J_i$ neighbors. In a process that does not affect the cyclic topology, it is assumed that, before the multiplex is assembled, the nodes are labeled differently in each individual layer. As the setup requires that all representations of the same node in the different $M$ layers must be linked by inter-layers connections, the resulting multiplex pattern can be highly complex. This can be better visualized if one assumes that in one layer the nodes are labeled in the sequentially ascending order and that the nodes with the same label are drawn on the top of the other, so that all inter-layer connections are represented by vertical lines. In such cases, the other $M-1$ layers are characterized by long distance shortcuts. As a consequence, in spite of sharing the same internal topological structure, the multiplex ends up with very dissimilar layers. For prime number of nodes, a regular star geometry arises by requiring that the neighbor labels of the $M-1$ layers differ by a constant value $p_i>1$. For $M=2$, we use analytical and numerical approaches to provide a thorough characterization of the multiplex topological properties, of the inter layer dissimilarity, and of the diffusive dynamical processes taking place on them. For the sake of definitiveness, it is considered that each node in the sequentially labeled layer is characterized by $J_1\geq1$. In the other layer, we fix $J_2\equiv1$, while $p>1$ becomes a proxy of layer dissimilarity.


\end{abstract}

\begin{titlepage}
\maketitle
\setcounter{page}{1}
\end{titlepage}

\section{Introduction}

In the last decade, the investigation of multilayer structures made evident their usefulness in characterizing several properties of actual real-world systems \cite{boccaletti14,kivela14,battiston17,aleta19}. By allowing for the description of more than one kind of interaction (embedded in distinct layers) among the systems agents described by the network nodes, they deeply impacted the study of structural and dynamical aspects of social \cite{szell10,cozzo13,li15,arruda17}, biochemical \cite{cozzo12,battiston17}, and transportation systems \cite{dedomenico14,aleta17}, among others. These include, for instance, clustering and community partition \cite{mucha10,boccaletti14,loe15,dedomenico15,zhai18,liu18}, synchronization \cite{gambuzza15,sevilla15,genio16,allen17}, diffusive behavior \cite{arruda18,gomez13,cencetti19,dedomenico16,tejedor18} or pattern formation \cite{asllani14,kouvaris15,busiello18}. Most of these studies focus on arrangements where the same nodes are present in all layers of interaction, denoted as multiplex networks \cite{cozzo_mult_18}.

One major issue in this theme is to understand how the properties of multilayer structures depend on the network dissimilarity among its individual layers \cite{brodka18,cencetti19,carpi19,serrano17}. In fact, conflicting or converging individual properties of each layer are expressed differently when they are joint together. The large diversity in the properties of single networks as well as in the possible ways to define assembling rules avoids the obtention of general results. Indeed, the current research scenario is still strongly devoted to identifying particular sets of networks and arrangements rules able to identify multilayers with particular properties observed in the analysis of actual systems.

We understand that a promising strategy to address this task is to start with topologically simple layers that allow to analytical expressions for the properties of multilayer networks. Among many advantages, this makes it easier to discern important measures to properly characterize the most important properties defining multilayer classes. Once such measures are well succeeded in identifying emerging geometrical properties of simple systems, they might well be suitable to analyze general networks.

Following these guidelines, in this work we investigate and characterize structural and dynamical properties of the class of prime star multiplex networks, where all layers are elements of a two-parameter sub-family of regular one dimensional cycles. Despite the apparently lack of flexibility offered by such configurations, we show how to assemble multiplexes with very different properties by adequately tuning the two free parameters. Also, the parameter choices are properly related to a measure of layer dissimilarity, making it possible to obtain a desired association between dissimilarity and multiplex properties.

In the first layer, the number of neighbors of any node defines the first parameter. The second layer, albeit consisting of a cycle with just two neighbors per node, stays under influence of the second parameter, which controls the choice of labels associated to the neighbors of each node. As multiplexes necessarily include inter-layer edges between the representation of a node in each pair of layers, the action of this parameter is sufficient to cause a strong dissimilarity between the layers, directly impacting the resulting multiplex properties. By imposing that the labels of neighboring nodes have a fixed difference, the simple cyclic topology remains preserved only if the number of nodes in the selected sub-family is prime.

The paper is organized as follows. In section \ref{Sec:concepts} we present the properties of individual layers, the different way of assembling the multiplex, and general formulation of diffusive dynamics in multiplexes. Section \ref{Sec:measures} discusses the relevant measures to quantitatively characterize the resulting structures and diffusive processes occurring on them. In section \ref{Sec:results} several analytical results are presented taking into account specific mathematical properties valid for the layers. They are complemented by several graphs obtained from the numerical evaluation of the derived expressions. Finally, our conclusions are summarized in section \ref{Sec:conclusions}.

\section{Prime stars multiplexes}
\label{Sec:concepts}

\subsection{Prime stars and interacting cycle graphs}

Let $G(V,E)$ be a simple (undirected and unweighted) graph without self-loops, where $V$ and $E$ represent, respectively, the set of nodes and the set of links. Let $N$ denote the number of nodes of $G$ and let $\mathbf{A}$ be the adjacency matrix of $G$, the elements of which are given by $\left ( \mathbf{A} \right )_{ij}=\left ( \mathbf{A} \right )_{ji}=1$ if nodes $i$ and $j\neq i$ are connected, and 0 otherwise. The degree of the node $i$ is given by $k_i=\sum_{m=1}^{N}=\left ( \mathbf{A} \right )_{im}$.

We denote a $N-$ring topology $G$ in which each node is connected to its $J$ left and $J$ right nearest nodes as \textit{interacting cycle graph}. Thus, when $J=1$, an interacting cycle is a \textit{cycle graph} and, for odd $N$ and $J=(N-1)/2$, it becomes a \textit{complete graph}. Interacting cycles are regular graphs, i.e., all nodes have the same degree $k\equiv k_i=2J$ for $i\in G$.

\begin{figure}[h!]
\centering
\includegraphics[width=0.51\textwidth]{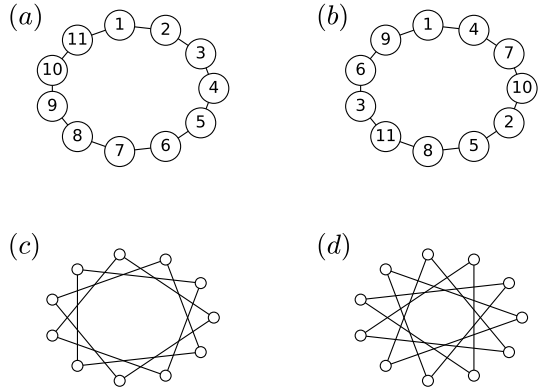}
\caption{Examples of prime stars with $N=11$ nodes. (a) $p=1$ (b) $p=3$ (c) $p=3$ (d) $p=4$. To display the nodes of prime stars in (a), (c) and (d) the same circular clockwise layout is used. However, for the sake of clarity, we only label the nodes in panels (a) and (b).}
\label{prime_examples}
\end{figure}

Let $1\leq i\leq N $ represent the label of a given node in $G$. In the case of cycle graphs, the nodes are usually labeled according to the simple rule illustrated in Fig.~\ref{prime_examples}a: node $i=1+j \left (\mathrm{mod}\,N  \right )$ is connected to nodes $i+1$ and $i-1$, where $"\mathrm{mod}"$ refers to the modulo operation and $0\leq j\leq N -1$.

For the purposes of this work, it is important to consider that, for a cycle graph with prime $N>2$ number of nodes, alternative choices of node labels can be used based on the following definition of the adjacency matrix: $\left ( \mathbf{A} \right )_{ij}=\left ( \mathbf{A} \right )_{ji}=1$ if $\left | i-j \right |=p$ or $N-p$, and 0 otherwise, for $1\leq p\leq (N-1)/2$. Fig.~\ref{prime_examples}b indicates that this rule is valid once, as required, it does not change the cycle graph topology. However, if one draws the nodes in the usual sequential clockwise order as in panel (a), the edges connecting the nodes with new labels give rise to regular geometrical figures identified as stars, as shown in Fig.~\ref{prime_examples}c and d. Therefore, throughout this work these arrangements will be identified as $(N,p)-$\textit{prime stars}. We remind that the $(N,p=1)-$ prime star corresponds to the usual labeling in Fig.~\ref{prime_examples}a.

The adjacency matrices for all interacting cycle graphs and for all $(N,p)-$ prime stars satisfy the following conditions:

\begin{enumerate}[label=(\roman*)]
\item They are circulant, i.e., all the elements $\left ( \mathbf{A} \right )_{ij}$ that meet the condition $\left | i-j \right |=m$ are equal, for $0\leq m\leq N-1$.
\item All the elements $\left ( \mathbf{A} \right )_{ij}$ with $j - i=m$ are equal to those with $j-i=N-m$, for $0\leq m\leq (N-1)/2$.
\end{enumerate}

\noindent Let us consider a generic $N\times N$ matrix $\mathbf{X}$ satisfying conditions (i) and (ii). Then, its elements are as follows:

\begin{align}
\mathbf{X} =
\left ( \begin{matrix}
x_0 & x_1 & x_2 & \cdots & x_3 & x_2 & x_1\\
x_1 & x_0 & x_1 & \cdots & x_4 & x_3 & x_2\\
x_2 & x_1 & x_0 & \cdots & x_5 & x_4 & x_3\\
\vdots & \vdots & \vdots & \ddots & \vdots  & \vdots &\vdots \\
x_3 & x_4 & x_5 & \cdots & x_0& x_1 & x_2\\
x_2 & x_3 & x_4 & \cdots & x_1 & x_0 & x_1\\
x_1 & x_2 & x_3 & \cdots & x_2 & x_1 & x_0
\end{matrix} \right ),
\label{circulant_s}
\end{align}

\noindent where, in general, $x_m\in\mathbb{C}$ for $0\leq m \leq (N-1)/2$. By definition, in the case of the adjacency matrices of interacting cycles and prime stars, $x_m\in\left \{ 0,1 \right \}$ for $m\neq0$ and $x_0=0$, and $\sum_{m=1}^{(N-1)/2}x_m=k=2J$. Consequently, for a given value of $J$, due to the condition $k=2J$, such adjacency matrices are well defined by only $(N-3)/2$ elements $x_m$ for $1\leq m \leq (N-1)/2$. For instance, the adjacency matrix of cycle graphs has elements $x_1=1$ and the remaining ones are equal to zero, whereas in the case of complete graphs all the elements of the adjacency matrix are equal to one, except $x_0=0$.

\subsection{Multiplexes}

Let us consider a multiplex $\mathcal{M}$ with $N$ nodes and $M$ layers. Let $\mathbf{A}_\alpha =\left ( \mathbf{A}_\alpha \right )_{ij}$ denote the adjacency matrix for the $\alpha$th layer with $1\leq \alpha \leq M$. In this work we focus on multiplexes $\mathcal{M}$ whose layers are undirected and unweighted, and contain no self-loops, i.e., $\left ( \mathbf{A}_\alpha \right )_{ij}=\left ( \mathbf{A}_\alpha \right )_{ji}=1$ if there is a link between the nodes $i$ and $j$ in the layer $\alpha$ (and $i\neq j$), and 0 otherwise.

Let us now define the $NM\times NM$ supra-adjacency matrix of $\mathcal{M}$ as follows

\begin{equation}
\mathbf{\mathcal{A}}^\mathcal{M}
=\left ( \begin{matrix}
\mathbf{A}^1 & \mathbf{I} & \cdots & \mathbf{I}\\
\mathbf{I} & \mathbf{A}^2 & \cdots & \mathbf{I}\\
 \vdots  & \vdots & \ddots  & \vdots \\
\mathbf{I} & \mathbf{I} & \cdots & \mathbf{A}^M
\end{matrix} \right ),
\label{def_sup_adj}
\end{equation}

\noindent where $\mathbf{I}$ represents the $N\times N$ identity matrix. This representation of the link structure of the multiplex is equivalent to a monolayer weighted network in which the node $f = i+(\lambda-1)N$ (with $i=1,\cdots,N$) describes the node $i$ at the $\lambda$th layer of $\mathcal{M}$. Besides that, it is easy to see that each node $i$ is linked to its counterparts in other layers.

\begin{figure}[h!]
\centering
\subfloat[]{
\centering
\includegraphics[width=0.5\linewidth]{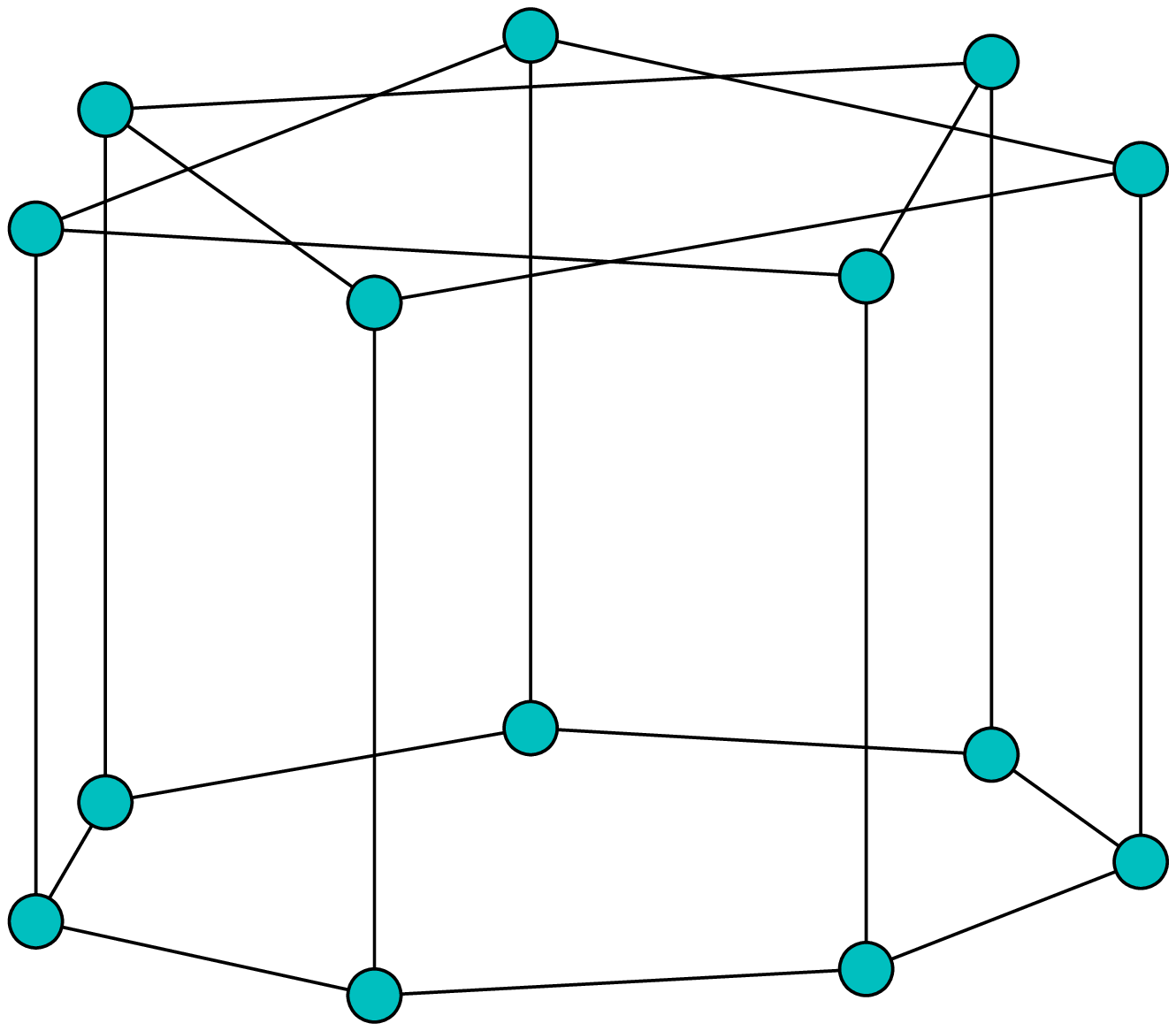}
\label{}
}
\subfloat[]{
\centering
\includegraphics[width=0.5\linewidth]{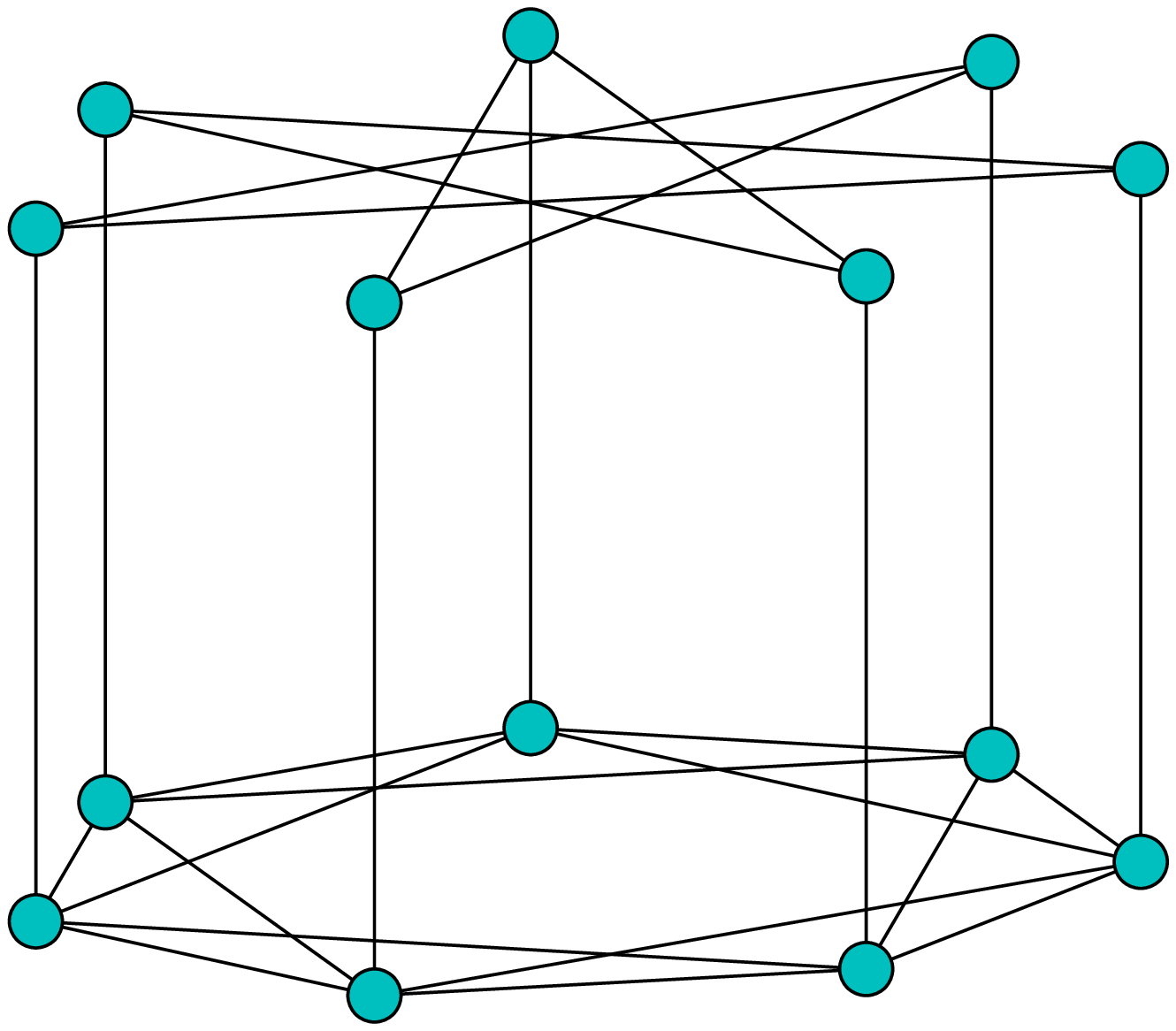}
\label{}
}
\caption{Examples of multiplex prime stars with $N=7$ nodes: $J=1$ and $p=2$ (a), and $J=2$ and $p=3$ (b).}
\label{exampl_PSM}
\end{figure}

Now let us consider a multiplex network $\mathcal{M}$ with $N$ nodes and $M$ layers such that at least one of them is a prime star with parameter $p_i>1$, and all the others are interacting cycle graphs with parameter $J_i$. We denote such systems as \textit{prime stars multiplexes}. For the sake of simplicity, in the rest of the work we restrict the discussion to the conditions $M=2$, $(J_1,p_1)=(1,p)$, and $(J_2,p_2)=(J,1)$. In Fig.~\ref{exampl_PSM} we show two examples of such arrangements, using the layout presented in Fig.~\ref{prime_examples}c and d to represent the prime star layers. This is the customary layout that puts a given node and all of its counterparts on top of the other, in such a way that all inter-layer connections are represented by vertical lines linking the different multiplex layers.

\subsection{Diffusion on multiplex}

Let $\vec{x}$ be a $NM\times1$ state (column) vector whose entry $i+(\alpha-1)N$ (with $i=1,\cdots,N$) describes the concentration of a generic flowing quantity at time $t$ on node $i$ at the $\alpha$th layer, $\left ( x_\alpha \right )_{i}$. The diffusion equation in matrix form reads

\begin{equation}
\frac{\mathrm{d}\vec{x}(t) }{\mathrm{d} t}=-\mathbf{\mathcal{L}}^\mathcal{M}\vec{x}(t),
\label{sol_prob}
\end{equation}

\noindent where

\begin{equation}
\mathbf{\mathcal{L}}^\mathcal{M}=\mathbf{\mathcal{L}}^\ell+\mathbf{\mathcal{L}}^x
\label{def_comb_lap}
\end{equation}

\noindent denotes the $NM\times NM$ combinatorial supra-Laplacian matrix defined in \cite{masuda17,gomez13,cencetti19,sole13} formed by two terms: $\mathcal{L}^\ell$ and $\mathcal{L}^x$ represent, respectively, the intra-layer and the inter-layer supra-Laplacian matrices expressed as

\begin{equation}
\mathcal{L}^\ell=\left ( \begin{matrix}
D_1 \mathbf{L}_1 &  &  & \\
 & D_2 \mathbf{L}_2 &  & \\
 &  & \ddots  & \\
 &  &  & D_M \mathbf{L}_M
\end{matrix} \right ),
\end{equation}

and

\begin{equation}
\mathcal{L}^x=\left ( \begin{matrix}
\sum_\alpha D_{1 \alpha}\mathbf{I} & -D_{1 2}\mathbf{I} & \cdots & -D_{1 M}\mathbf{I}\\
-D_{21}\mathbf{I} & \sum_\alpha D_{2 \alpha}\mathbf{I} & \cdots & -D_{2M}\mathbf{I}\\
\vdots  & \vdots & \ddots  & \vdots \\
-D_{M 1}\mathbf{I} & -D_{M 2}\mathbf{I} & \cdots & \sum_\alpha D_{M \alpha}\mathbf{I}
\end{matrix} \right ).
\label{interlayer_connect_matrix}
\end{equation}

The diagonal blocks of $\mathcal{L}^\ell$ are given by the product of the intra-layer diffusion constant in the $\alpha$th layer $D_\alpha$ and $\mathbf{L}_\alpha$, which is the usual $N\times N$ Laplacian matrix of the layer $\alpha$. Its matrix elements are $\left ( \mathbf{L}_\alpha \right )_{ij}=\left ( k_\alpha \right )_{i}\delta_{ij} -  \left ( \mathbf{A}_\alpha \right )_{ij}$, where $\left ( k_\alpha \right )_{i}$ represents the degree of node $i$ with respect to its connections with other vertices in the same layer $\alpha$, and $\delta_{ij}$ is the Kronecker delta function. In $\mathcal{L}^x$, $D_{\alpha\beta}$ (with $\alpha,\beta\in\left \{ 1,\cdots,M \right \}$ and $\beta\neq \alpha$) indicates the inter-layer diffusion constant between the $\alpha$th and $\beta$th layers, and $\mathbf{I}$ corresponds to the $N\times N$ identity matrix defined in (\ref{def_sup_adj}). For the sake of simplicity, we will consider only diffusion processes where $D_{\alpha\beta}=D_{\beta\alpha}$, so that $\mathcal{L}^x$ and $\mathbf{\mathcal{L}}^\mathcal{M}$ are symmetric matrices. We remark that our results for multiplex diffusion are based on the edge-centric description of the process. Other node-centric processes, which are not based on the use of the combinatorial supra-Laplacian as defined by Eqs.~(\ref{def_comb_lap}-\ref{interlayer_connect_matrix}), will not be considered.

Interestingly, due to the circulant property of the adjacency matrices $\mathbf{A}$ of interacting cycles and prime stars, their respective Laplacian matrices $\mathbf{L}$ as well as the corresponding blocks of the supra-adjacency and supra-Laplacian matrices $\mathbf{\mathcal{A}}^\mathcal{M}$ and $\mathbf{\mathcal{L}}^\mathcal{M}$ associated to the prime-stars multiplexes also satisfy (\ref{circulant_s}).

\section{Relevant measures}
\label{Sec:measures}

To characterize structural and diffusive properties of the multiplexes investigated in this work, in this section we present a brief description of the following relevant quantitative measures: edge overlap, network dissimilarity, and algebraic connectivity.

Starting with a formal definitions related with graph theory, a \textit{path} of length $\ell$ in $G$ is a set of nodes $i_1$, $i_2$, $\cdots$, $i_{\ell}$, $i_{\ell+1}$ such that for all $1\leq r \leq \ell$, $(i_r,i_{r+1}) \in E$ with no repeated nodes. We consider that there is a path between every pair of vertices $i,j\in V$. The \textit{shortest-path} $d_G(i,j)$ is defined as the length of the shortest path connecting the nodes $i$ and $j$. The network diameter is defined as $\mathfrak{D}_G = \max_{i,j} d_G(i,j)$.

Given any two networks $\alpha$ and $\beta$ with the same number of
nodes $N$, it is also possible to define the fraction of connected pairs $i$ and $j$ that are linked in both networks, i.e., the \textit{edge overlap}, denoted by $w$. Following \cite{cencetti19}, we measure $w$ as follows:

\begin{align}
w=\frac{\sum_{i=1}^N \sum_{j>i}^N \left ( \mathbf{A}_\alpha  \right )_{ij}\left ( \mathbf{A}_\beta  \right )_{ij}}{\sum_{i=1}^N \sum_{j>i}^N \left ( \left ( \mathbf{A}_\alpha  \right )_{ij}+\left ( \mathbf{A}_\beta  \right )_{ij}-\left ( \mathbf{A}_\alpha  \right )_{ij}\left ( \mathbf{A}_\beta  \right )_{ij} \right )},
\label{edge_w}
\end{align}

\noindent where $\mathbf{A}_\alpha$ and $\mathbf{A}_\beta$ are the corresponding adjacency matrices of $\alpha$ and $\beta$.

For the purposes of this work, we also use a measure of \textit{network dissimilarity} to characterize how different the multiplex layers are. Here we adopt dissimilarity measure provided by the topological distance $\delta$, defined for any two networks $\alpha$ and $\beta$ with the same number of nodes $N$ as \cite{serrano17,Andrade08}:

\begin{align}
\delta^2= \frac{1}{N(N-1)}\sum_{i=1}^N \sum_{j=1}^N\left ( \frac{\left ( \widehat{\mathbf{V}}_\alpha \right )_{ij}}{\mathfrak{D}_\alpha} -\frac{\left ( \widehat{\mathbf{V}}_\beta \right )_{ij}}{\mathfrak{D}_\beta} \right )^2.
\label{distance_def}
\end{align}

\noindent Here, $\widehat{\mathbf{V}}_\alpha$ and $\widehat{\mathbf{V}}_\beta$ represent the neighborhood matrices of each network \cite{Andrade08,Andrade06}, the elements of which are defined by $\left ( \widehat{\mathbf{V}}_G \right )_{ij}=d_{G}(i,j)$. As expected, the larger the distance $\delta$, the more different the elements are. We observe that, although many other dissimilarity measures have been introduced in the literature (see for instance, the definition of layer difference presented in \cite{carpi19}), the features captured by the adopted function $\delta$ are relevant for our work. We also remind that, if $\mathbf{A}_\alpha$ displays the circulant structure (\ref{circulant_s}), the same is also observed for the corresponding neighborhood matrix $\widehat{\mathbf{V}}_\alpha$.

The \textit{algebraic connectivity} of the supra-Laplacian matrix $\mathbf{\mathcal{L}}^\mathcal{M}$ is defined as its second-smallest eigenvalue and denoted here as $\Lambda_2$ \cite{note1}. It characterizes the time of convergence of the diffusion process $\tau^\mathcal{M}$, which is given by $\tau^\mathcal{M}\sim 1/\Lambda_2$ \cite{gomez13,cencetti19,tejedor18}. To emphasize the inter-layer diffusion process and simplify the notation, we choose the diffusion coefficients $D_1=D_2=1$ and $D_{12}/D_\alpha=D_x$ for $\alpha \in \left \{ 1, 2 \right \}$ \cite{gomez13,cencetti19,tejedor18}.

\section{Results}
\label{Sec:results}

\subsection{Edge overlap and layer dissimilarity}

\begin{figure}[h!]
\centering
\subfloat[]{
\centering
\includegraphics[width=0.5\linewidth]{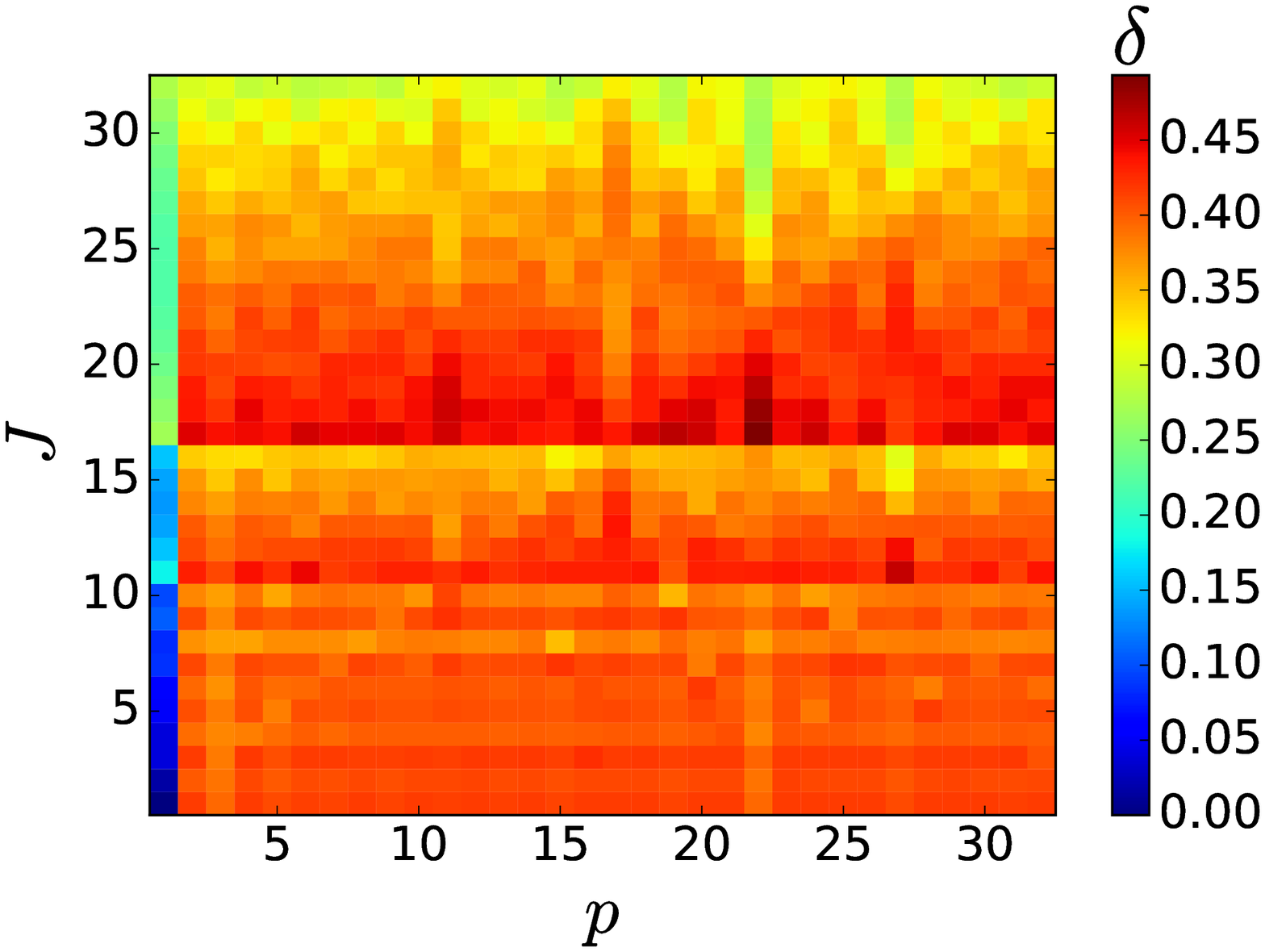}
\label{}
}
\subfloat[]{
\centering
\includegraphics[width=0.5\linewidth]{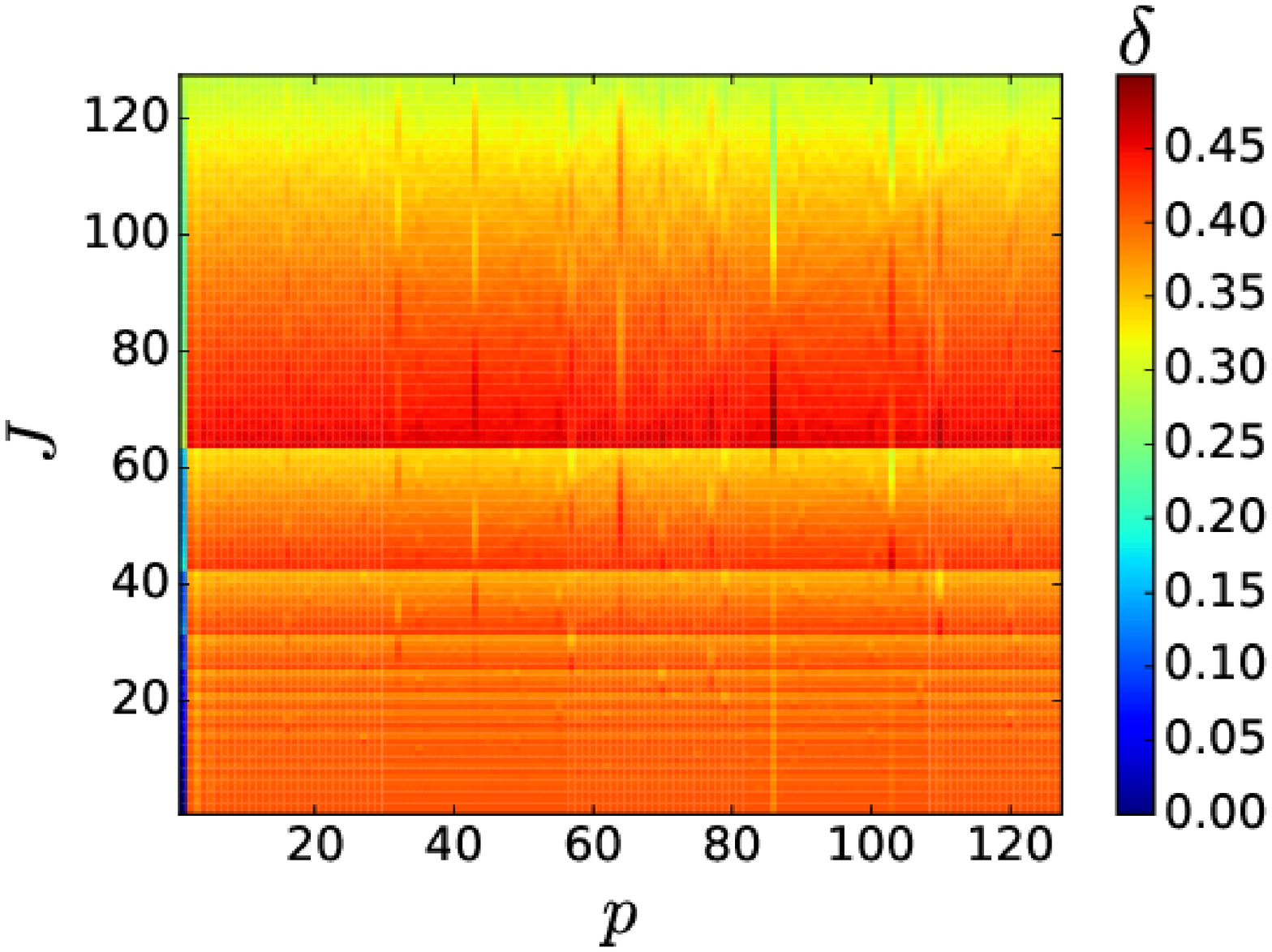}
\label{}
}
\caption{Dependence of the network dissimilarity $\delta$ on $J$ and $p$ for interacting cycles and prime stars with $N=67$ (a) and $N=257$ nodes (b).}
\label{delta_Jp}
\end{figure}

We start the section presenting the measures of edge overlap, taking into account of the circulant property (\ref{circulant_s}). After conducting the necessary manipulations, in the case of an interacting cycle graph and a prime star it is possible to reduce Eq.~(\ref{edge_w}) to the following expression

\begin{align}
w=\left\{\begin{matrix}
1/J & \mathrm{for}\;p\leq J\\
0 & \mathrm{otherwise}
\end{matrix}\right..
\label{edge_w_interac}
\end{align}

\noindent This result calls the attention to two geometrical properties, which can also be be easily identified by inspection: no edge overlap exists if $p>J$ and, provided this condition is not satisfied, the fraction of overlapping edges depends only on $J$, not on $p$. Also, it shows that interacting cycles and prime stars are equivalent (i.e., $w=1$) only when $Jp=1$.

Regarding the evaluation of layer dissimilarity, we have to consider in Eq.~(\ref{distance_def}) that $\alpha$ and $\beta$ represent, respectively, an interacting cycle graph and a prime star. In the case of $\alpha$, the elements of $\widehat{\mathbf{V}}_\alpha$ are defined by $v_m^\alpha\equiv x_m=\left \lceil m/J \right \rceil$ for $1\leq m \leq (N-1)/2$, where $\left \lceil . \right \rceil$ is the ceil function. In the case of $\beta$, the elements of $\widehat{\mathbf{V}}_\alpha$ are determined by $v_m^\beta\equiv x_{m}= d$ for

\begin{align}
m=\left\{\begin{matrix}
\min \left ( j, N+2-j \right ) - 1 & \mathrm{for}\;dp+1 \,\left (\mathrm{mod}\,N  \right )\neq 0\\
1 & \mathrm{otherwise}
\end{matrix}\right.,
\label{neigh_prime}
\end{align}

\noindent where $j=dp+1 \,\left (\mathrm{mod}\,N  \right )$, $1\leq d \leq (N-1)/2$, and $\min$ denotes the minimum value of a set of elements. As expected, when $p=1$ we obtain $v_m^\beta=m$ from (\ref{neigh_prime}). On the other hand, if $N>2$ is a prime number, the diameters of $\alpha$ and $\beta$ are given respectively by $\mathfrak{D}_\alpha=\left \lceil (N-1)/(2J) \right \rceil$ and $\mathfrak{D}_\beta=(N-1)/2$. Therefore, according to (\ref{distance_def}), the adopted dissimilarity measure between an interacting cycle graph and a prime star layers are expressed by

\begin{align}
\delta^2= \frac{2}{N-1}\sum_{m=1}^{(N-1)/2}\left ( \frac{\left \lceil m/J \right \rceil}{\left \lceil (N-1)/(2J) \right \rceil} -\frac{2v_m^\beta}{N-1} \right )^2,
\label{distance_def_reduced}
\end{align}

\noindent for $Jp\neq1$, and zero otherwise. In Fig.~\ref{delta_Jp} we show two examples of the dependence of the network dissimilarity $\delta$ on $J$ and $p$ for interacting cycles and prime stars.

A first feature to be noticed is that, independently of the value of $N$, $\delta$ occur in different ranges when $p>1$ ($ 0.28 \lesssim  \delta \lesssim 0.5$) and $p=1$ ($0 \leq \delta \lesssim 0.28 $). In this last case, the condition $\delta=0$ is valid only for two identical layers corresponding to $J=1$. Next we observe the presence of several horizontal stripes, characterized by $\delta \approx 0.45$ on the bottom and decreasing to $\delta \approx 0.3$ on the top. The values of $J$ at which the limits of each stripe can be identified are $J\simeq (N-1)/2\ell$, for some integer values of $\ell$. As $J$ and $p$ are integers, the number of identifiable stripes increase with respect to $N$. Superimposed to the dominating horizontal pattern, we identify several vertical lines characterized by distinct values of $\delta$ which, depending on the range of $J$, can be larger or smaller than those at their immediate neighborhood. Both the horizontal and vertical discontinuities are directly related to the presence of the modulo operation and of the ceil function in Eq.~(\ref{distance_def_reduced}), while the specific dependences also explain why, for a given $N$ and $J$, it is possible to obtain the same value of $\delta$ for different values of $p$. For instance, taking $N=11$ and $J=1$, $\delta(N,J,p=3)=\delta(N,J,p=4)\approx 0.44$. Our results in the last part of this section show that the vertical stripes at specific values of $p$ also influence the algebraic connectivity of the multiplexes.

The comparison between figures \ref{delta_Jp}a and b suggest the scale invariance of the resulting patterns with respect to the number of nodes. Although they become clearer for larger values of $N$, we use smaller values of $N$ to identify some specific aspects that become more visible in a coarse grained view. Thus it is possible to realize that, given two system sizes $N_1$ and $N_2\neq N_1$, respectively, the corresponding values of $\delta$ for interacting cycles and prime stars with size $N_\alpha$ and parameter combinations $(J_\alpha,p_\alpha)$ (where $\alpha\in\left \{ 1,2 \right \}$ and $J_\alpha,p_\alpha \in\mathbb{N}$) are approximately equal if (i) $J_1\approx J_2 (N_1-1)/(N_2-1)\approx J_2 N_1/N_2$ and (ii) $p_1\approx p_2(N_1-1)/(N_2-1)\approx p_2 N_1/N_2$ (note that, since $N_1$ and $N_2$ are prime numbers, $N_1/N_2\notin \mathbb{Z}$). For instance, it is possible to identify in both panels the diagonal line at $J_\alpha=p_\alpha$, the vertical stripes at $p_\alpha=1$ and $p_\alpha\approx(N_\alpha-1)/3$, and the horizontal line at $J_\alpha\approx(N_\alpha-1)/4$, among many others.

\subsection{Algebraic connectivity}
\label{Subsec:algebraic_conn_mult}

\begin{figure}[h!]
\centering
\subfloat[]{
\centering
\includegraphics[width=0.5\linewidth]{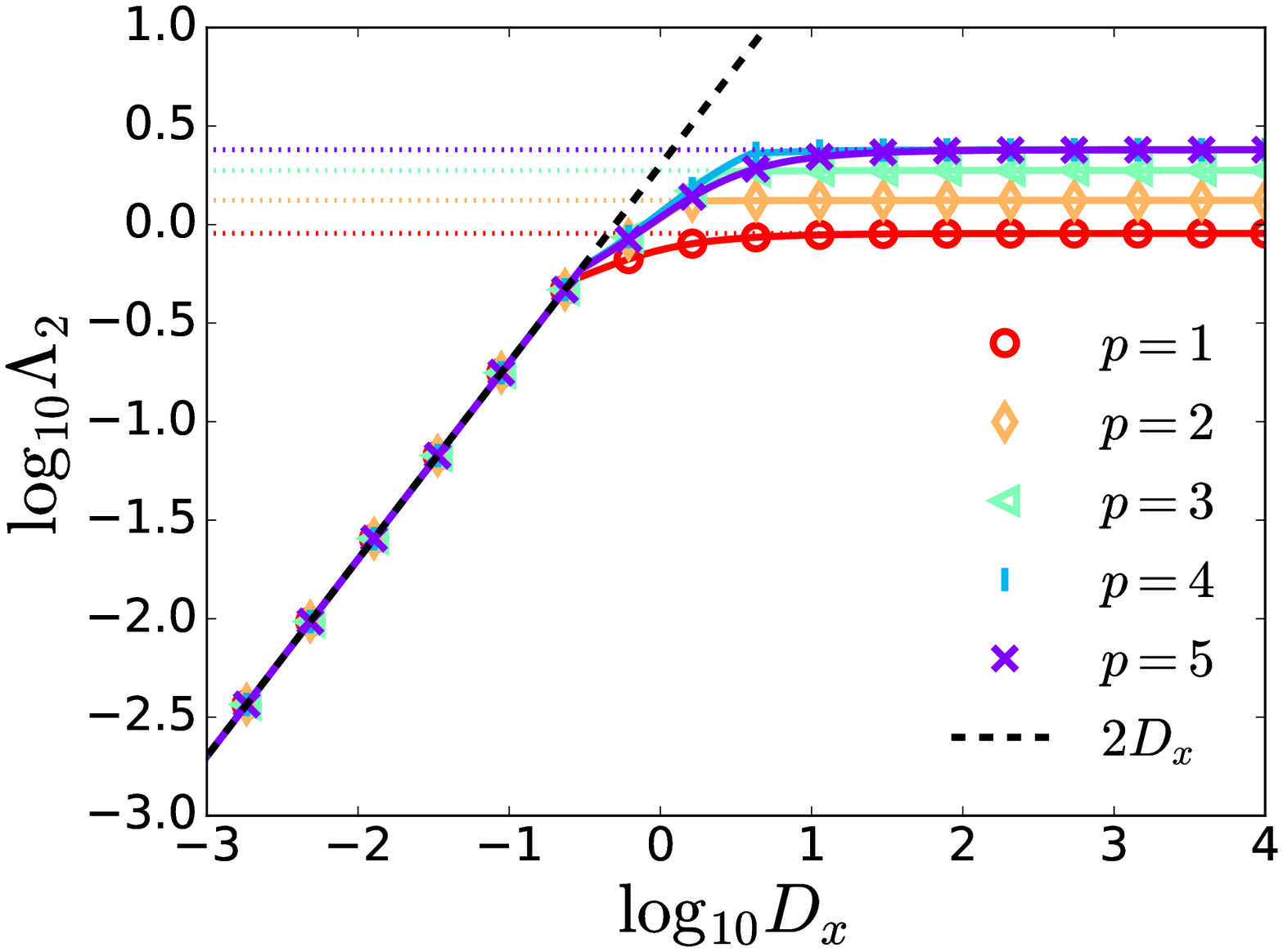}
\label{}
}
\subfloat[]{
\centering
\includegraphics[width=0.5\linewidth]{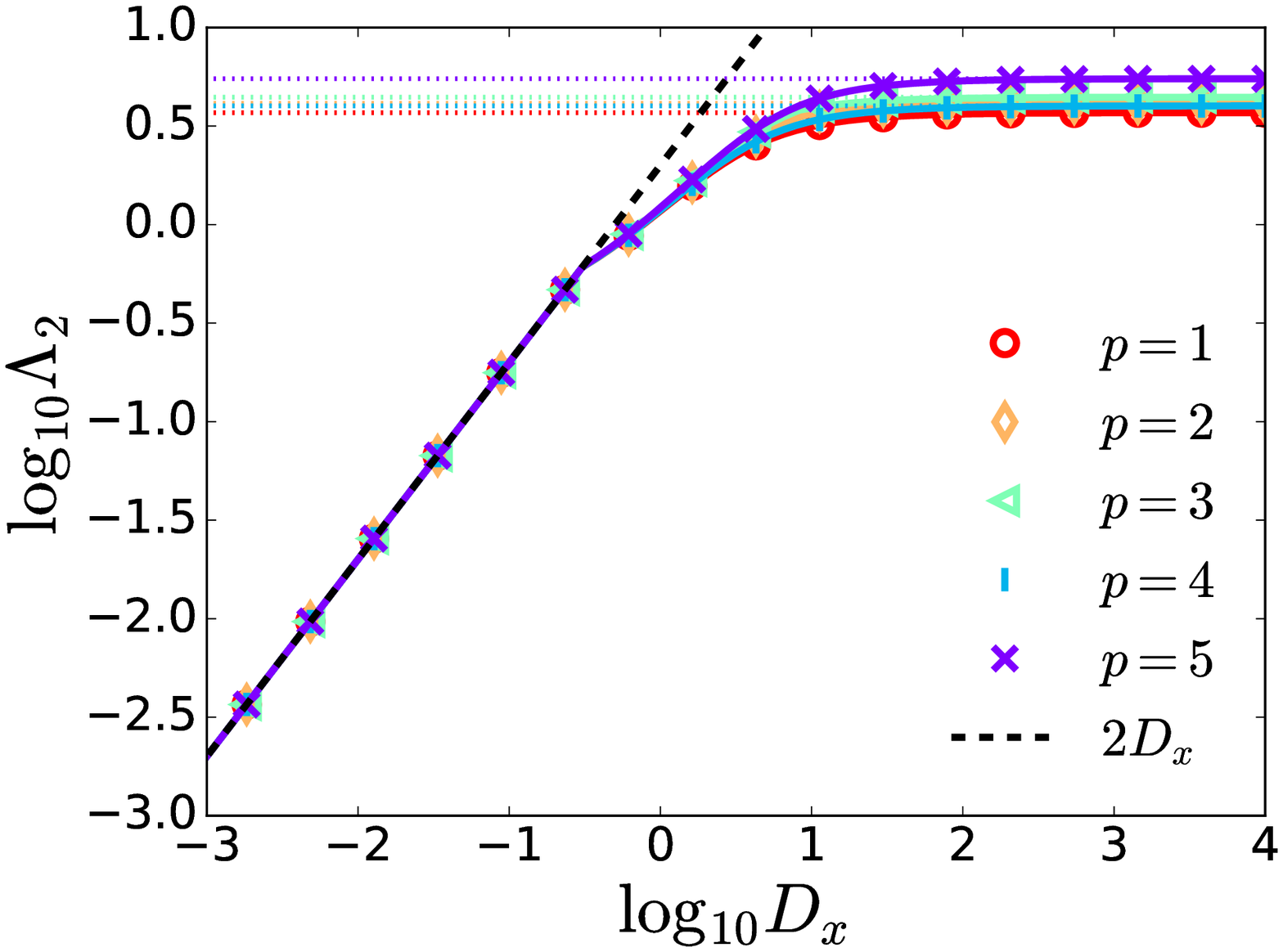}
\label{}
}
\caption{Dependence of $\Lambda_2$ on $D_x$ for prime star multiplex networks with $N=11$ nodes, when $J=2$ (a) and $J=4$ (b): $p=1$ (circles), $p=2$ (diamonds), $p=3$ (triangles), $p=4$ (bars), and $p=5$ (crosses). Black dashed line represents $2D_x$, whereas dotted lines show the algebraic connectivity of $\bar{\mathbf{L}}$ for each $(J,p)-$combination.}
\label{Example_alge_dep_Dx}
\end{figure}

Here we treat analytically the algebraic connectivity of the supra-Laplacian matrix, taking into account that the blocks of $\mathbf{\mathcal{L}}^\mathcal{M}$ are circulant and making use of general analytical expressions for the eigenvalues and eigenvectors of such matrices \cite{Mieghem11}. For the conditions stated above, and considering the definition $\mathbf{C}_\alpha=\mathbf{L}_\alpha +D_x\mathbf{I}$, it is possible to write down the general expression

\begin{equation}
\mathbf{F}^{-1}
\left ( \begin{matrix}\mathbf{C}_1 & -D_x\mathbf{I}\\
-D_x\mathbf{I} & \mathbf{C}_2
\end{matrix} \right ) \mathbf{F}
=
\left ( \begin{matrix}\mathrm{\Xi}_1 & -D_x\mathbf{I}\\
-D_x\mathbf{I} & \mathrm{\Xi}_2
\end{matrix} \right ),
\end{equation}

\noindent where $\mathbf{F}$ is a $2N\times 2N$ block-diagonal matrix formed by two identical $N\times N$ hermitian $\mathbf{U}$ matrices, with elements $\left ( \mathbf{U} \right )_{ij}=\omega ^{(i-1)(j-1)}/\sqrt{N}$, $\omega \equiv \exp(-\mathfrak{i}2\pi/N)$, and $\mathfrak{i}=\sqrt{-1}$. On the other hand, $\mathrm{\Xi}_\alpha=\mathrm{diag}\left (  \xi_{1,\alpha} ,\cdots, \xi_{{N},\alpha}  \right )$, and $\xi_{m,\alpha}$ are the eigenvalues of $\mathbf{C}_\alpha$. In the case of interacting cycles, the eigenvalues of $C_1$ are

\begin{align}
\xi_{m,1}&=D_x+A_{m}^1=D_x+2\left (J+1\right )
-2\frac{\sin\left ((J +1)\frac{\pi(m-1)}{N}\right )\cos\left (J\frac{\pi(m-1)}{N}\right )}{\sin\left (\frac{\pi(m-1)}{N}\right )}
\label{eigen_xi_interacting}
\end{align}

\noindent for $1<m\leq N$, and $\xi_{m,1}=D_x$ for $m=1$. On the other hand, in the case of prime stars, the eigenvalues of $C_2$ are

\begin{align}
\xi_{m,2}=D_x+A_m^2=D_x+2-2\cos\left (p\frac{2\pi(m-1)}{N}\right ),
\label{eigen_xi_prime}
\end{align}

\noindent for $1\leq m\leq N$. Since the matrices $-D_x\mathbf{I}$ and $\left (\mathrm{\Xi}_2-\mu_m\mathbf{I}\right )$ conmute, the eigenvalues of $\mathbf{\mathcal{L}}^\mathcal{M}$ can be obtained as:

\begin{equation}
\sigma_{2m-1}= \frac{\xi_{m,1}+\xi_{m,2}+\sqrt{\left (\xi_{m,1}-\xi_{m,2}\right )^2+4D_x^2}}{2},
\label{lapl_eigenA}
\end{equation}

\noindent and

\begin{equation}
\sigma_{2m}=\frac{\xi_{m,1}+\xi_{m,2}-\sqrt{\left (\xi_{m,1}-\xi_{m,2}\right )^2+4D_x^2}}{2},
\label{lapl_eigenB}
\end{equation}

\noindent for $m\in\left \{ 1,\cdots,N \right \}$. Note that the eigenvalues $\sigma_m$ are not ordered from smallest to largest and vice versa (for instance, when $m=1$, $\sigma_2=0$). According to (\ref{lapl_eigenA}) and (\ref{lapl_eigenB}), $\Lambda_2$ depends on the values of $D_x$, $N$, $J$ and $p$. When $D_x\ll 1$, $\Lambda_2 = \sigma_{1}=2D_x$, as previously reported in the literature (see Fig.~\ref{Example_alge_dep_Dx} for an example). On the other hand, when $D_x\gg 1$, it is possible to see that  $\Lambda_2 = \sigma_{2c} < \sigma_{2c-1}$ where $c\in \left \{ 2,\cdots,N \right \}$ is the natural number that minimizes $\sigma_{2c}$. As expected, in the limit of $D_x\rightarrow \infty$, this value of $\Lambda_2$ converges to the algebraic connectivity of the Laplacian matrix $\bar{\mathbf{L}} = (\mathbf{L}_1 + \mathbf{L}_2)/2$, which represents the superposition of the two layers. These aspects are shown Fig.~\ref{Example_alge_dep_Dx} for different parameter choices. For the sake of a better visualization, we use a small value $N=11$. It is also possible to see that, when $D_x\gg 1$, given a $(N,p)-$combination, the larger the value of $J$, the larger $\Lambda_2$. Adding extra connections to the interacting cycle graph speeds up the diffusion in that layer, and, consequently, it reduces the time of convergence of the multiplex diffusion process $\tau^\mathcal{M} \sim 1/\Lambda_2$.

\begin{figure}[h!]
\centering
\subfloat[]{
\centering
\includegraphics[width=0.50\linewidth]{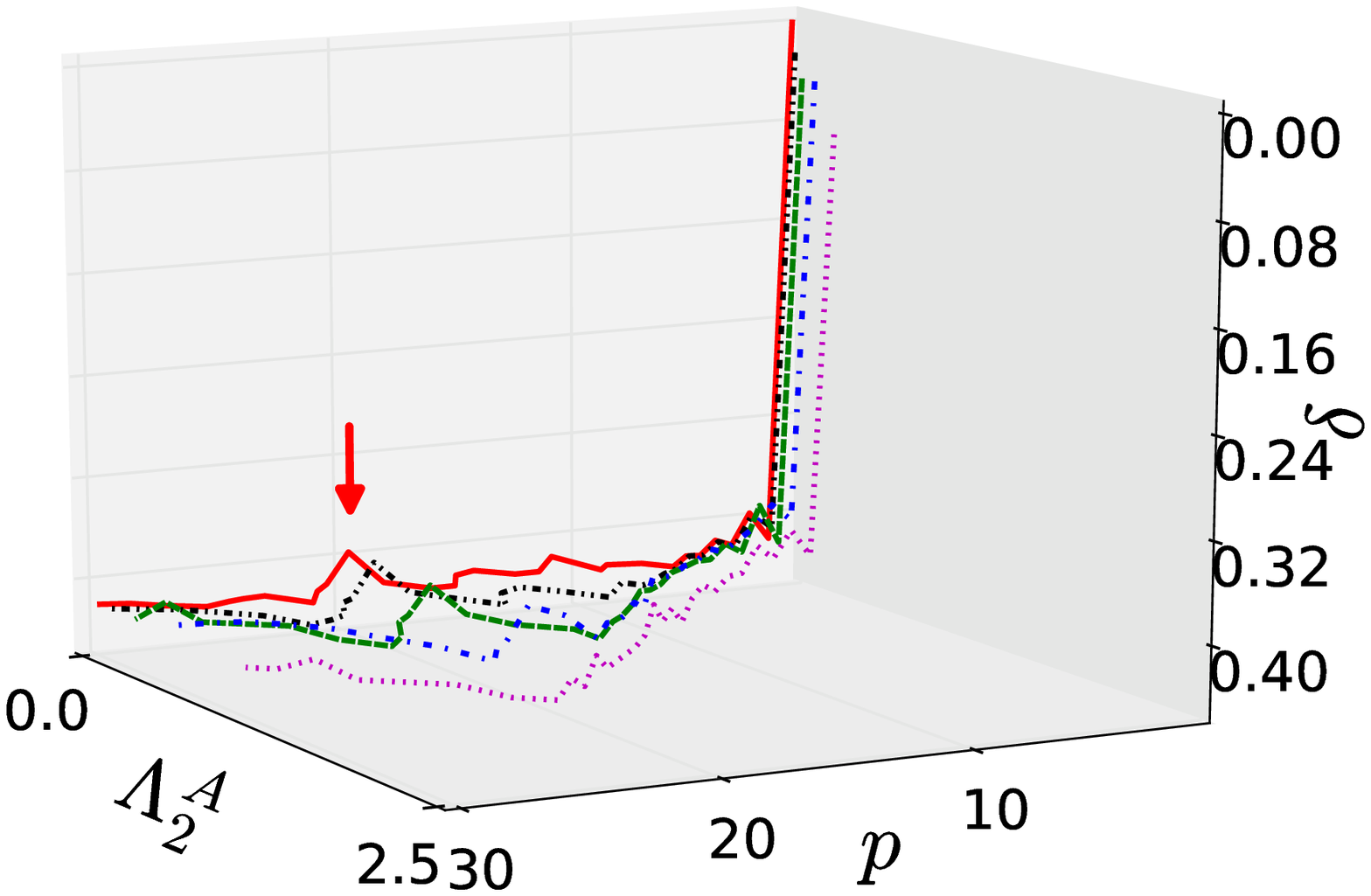}
\label{}
}
\subfloat[]{
\centering
\includegraphics[width=0.50\linewidth]{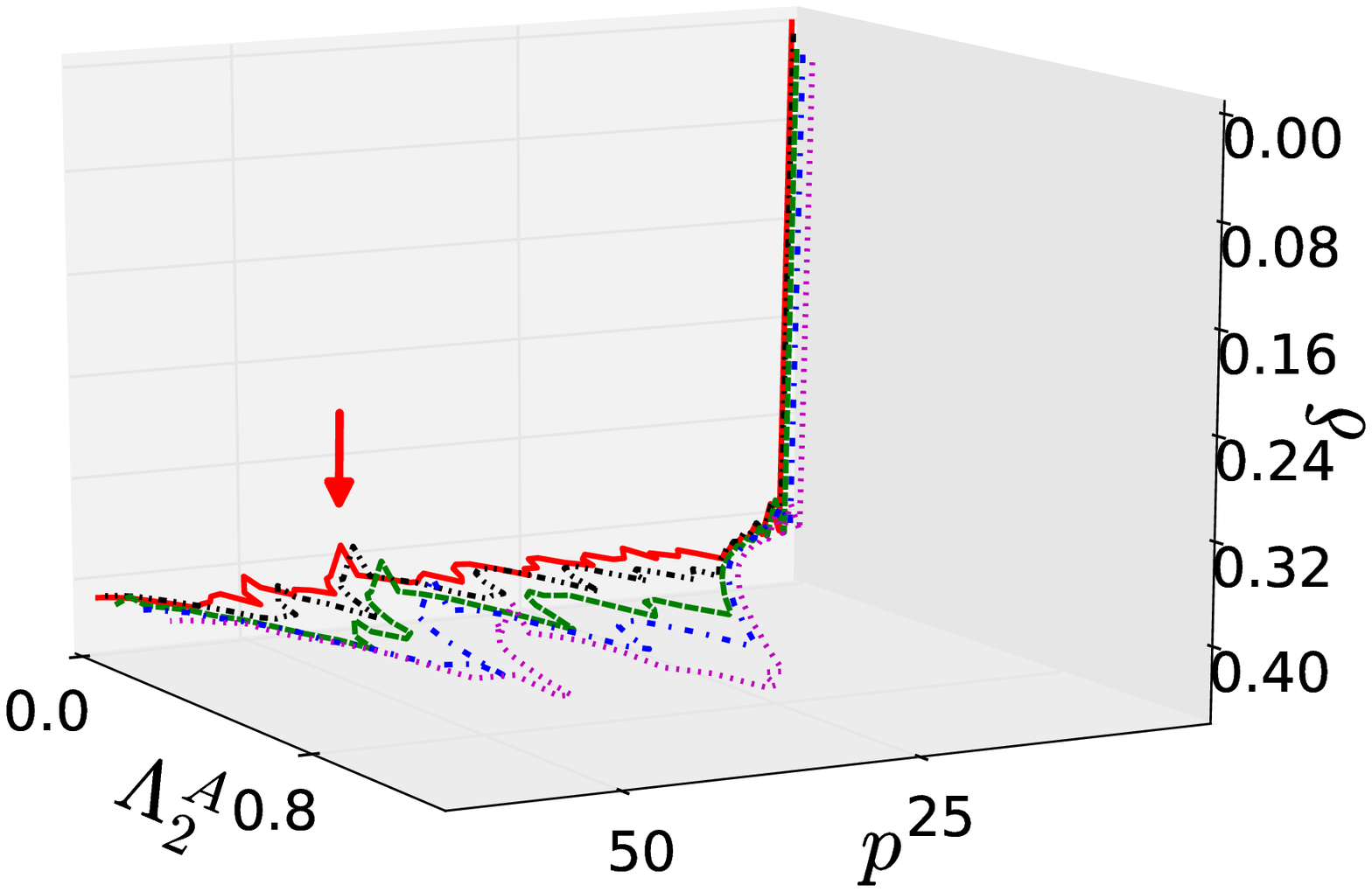}
\label{}
}
\caption{Dependence of $\Lambda_2^A$ on $J$, $p$ and $\delta$ for prime stars multiplex with $N=61$ (a) and $N=127$ nodes (b): $J=1$ (red solid line), $J=2$ (black dash-dot-dotted line), $J=3$ (green dashed line), $J=4$ (blue dash-dotted line), and $J=5$ (magenta dotted line). Red arrows locate the relative minimum value of $\delta$ at $p\approx(N-1)/3$, when $J\sim1$.}
\label{Lambda_2_fig}
\end{figure}

To better appreciate the dependence of $\Lambda_2$ for large values of $N$, it is convenient to consider the asymptotic values of $\Lambda_2$ obtained in the limit of large interlayer coupling. When $D_x\rightarrow \infty$, $\left (\xi_{m,1}-\xi_{m,2}\right )^2+4D_x^2 \rightarrow 4D_x^2$, and it is possible to write

\begin{align}
\Lambda_2^A(N,J,p)&\equiv \lim_{D_x\rightarrow \infty}\Lambda_2 = \lim_{D_x\rightarrow \infty}\lambda_{2c}=\frac{A_c^1+A_c^2}{2}\nonumber\\
&=2+J-\cos\left (p\frac{2\pi(c-1)}{N}\right )
-\frac{\sin\left ((J +1)\frac{\pi(c-1)}{N}\right )\cos\left (J\frac{\pi(c-1)}{N}\right )}{\sin\left (\frac{\pi(c-1)}{N}\right )},
\label{asin_eigenB}
\end{align}

\noindent where $c\in \left \{ 2,\cdots,N \right \}$ is the natural number that minimizes (\ref{asin_eigenB}). As an example, in Fig.~\ref{Lambda_2_fig} we show the dependence of $\Lambda_2^A$ on $J$ and $p$ for $N=61$ and $N=127$. As can be observed, for any given value of $p$ and $N$, if $J_1<J_2$, $\Lambda_2^A(N,J_1,p) < \Lambda_2^A(N,J_2,p)$. On the other hand, for a given $N$ and $J$, the monotonic dependence of $\Lambda_2^A(N,J,p)$ on $p$ is not is not guaranteed. Indeed, when $J=(N-1)/2$ (i.e., when the interacting cycle is a complete graph), for any given $p$, the value of $\Lambda_2^A(N,(N-1)/2,p)=1+(N/2)-\cos(2\pi/N)$ is constant.

In Fig.~\ref{Lambda_2_fig} it is also possible to follow the dependence between $\Lambda_2^A$ and $\delta$. The obtained curves for any chosen value of $J$ clearly indicate a complex, non-monotonic behavior among $\Lambda_2^A$, $p$ and $\delta$. Indeed, once the interacting cycle's topology is defined, increasing the network dissimilarity between the layers of the prime star multiplex does not guarantee that $\Lambda_2^A$ will also increase. As can be observed even for the simplest case $J=1$, in which both layers correspond to a cycle graph, an increase in $\delta$ does not necessarily lead to an increase in $\Lambda_2^A$. This non-monotonic trend was previously observed in \cite{serrano17}, when analyzing small-world (Watts-Strogatz) networks \cite{watts98}. To conclude, it is worth mentioning that the layers distance $\delta$ exhibits a relative minimum at $p\approx(N-1)/3$, when $J\sim1$ (see Fig.~\ref{Lambda_2_fig}).


\subsection{Structural super-diffusion}
\label{Subsec:superdiffusion}

In this subsection we study the topological conditions that allow the emergence of \textit{structural super-diffusion} on prime star multixplex networks, i.e., the combinations of $J$ and $p$ (for a given $N$) that make the diffusion time scale of the system smaller than that of each layer. To this purpose, we first let $\lambda_2^\alpha$ be the algebraic connectivity of the $\alpha-$th layer of the multiplex network in isolation (with $\alpha\in\left \{ 1,\cdots,M \right \}$), and $\tau^\alpha$ be the diffusion time scale of such isolated layer, where $\tau^\alpha\sim1/\lambda_2^\alpha$ \cite{gomez13}. As shown in \cite{gomez13}, in many situations $\tau^\mathcal{M}$ is very different from $\tau^\alpha$. The particular condition $\tau^\mathcal{M}<\tau^\alpha$ for $\alpha\in\left \{ 1,\cdots,M \right \}$ has great importance for many systems, and is referred to as (structural) super-diffusion \cite{gomez13}.

In the case of prime stars multiplexes, it is possible to calculate analytically the algebraic connectivity of their layers. We denote the algebraic connectivity of interacting cycle graphs and that of prime stars as $\lambda_2^1$ and $\lambda_2^2$, respectively. For interacting cycles, the algebraic connectivity reads

\begin{align}
\lambda_2^1&=2\left (J+1\right ) -2\frac{\sin\left ((J +1)\frac{\pi}{N}\right )\cos\left (J\frac{\pi}{N}\right )}{\sin\left (\frac{\pi}{N}\right )}=2J+1-\frac{\sin\left (\frac{\pi}{N}(2J +1)\right )}{\sin\left (\frac{\pi}{N}\right )},
\label{algebraic_interacting}
\end{align}

\noindent whereas for prime stars it can be written as

\begin{align}
\lambda_2^2=2-2\cos\left (p\frac{2\pi(m_c-1)}{N}\right ),
\label{algebraic_prime}
\end{align}

\noindent where $m_c\in \left \{ 2,\cdots,N \right \}$ is the natural number that meets the following condition: $p(m_c-1)$ mod $N = 1$. To identify super-diffusive configurations, we compare the algebraic connectivity of the prime star multiplex with that of the fastest layer. For that reason, we define the following absolute indicator

\begin{align}
\eta=\frac{\Lambda_2}{\max_\alpha\left ( \lambda_2 ^\alpha \right ), }
\label{def_superdiff}
\end{align}

\noindent in such a way that super-diffusion emerges whenever $\eta>1$. We further observe that, for any $J>1$, $\max_\alpha\left ( \lambda_2 ^\alpha \right ) = \lambda_2^1$, due to the larger amount of connections of the interacting cycle graph.

\begin{figure}[h!]
\centering
\subfloat[]{
\centering
\includegraphics[width=0.5\linewidth]{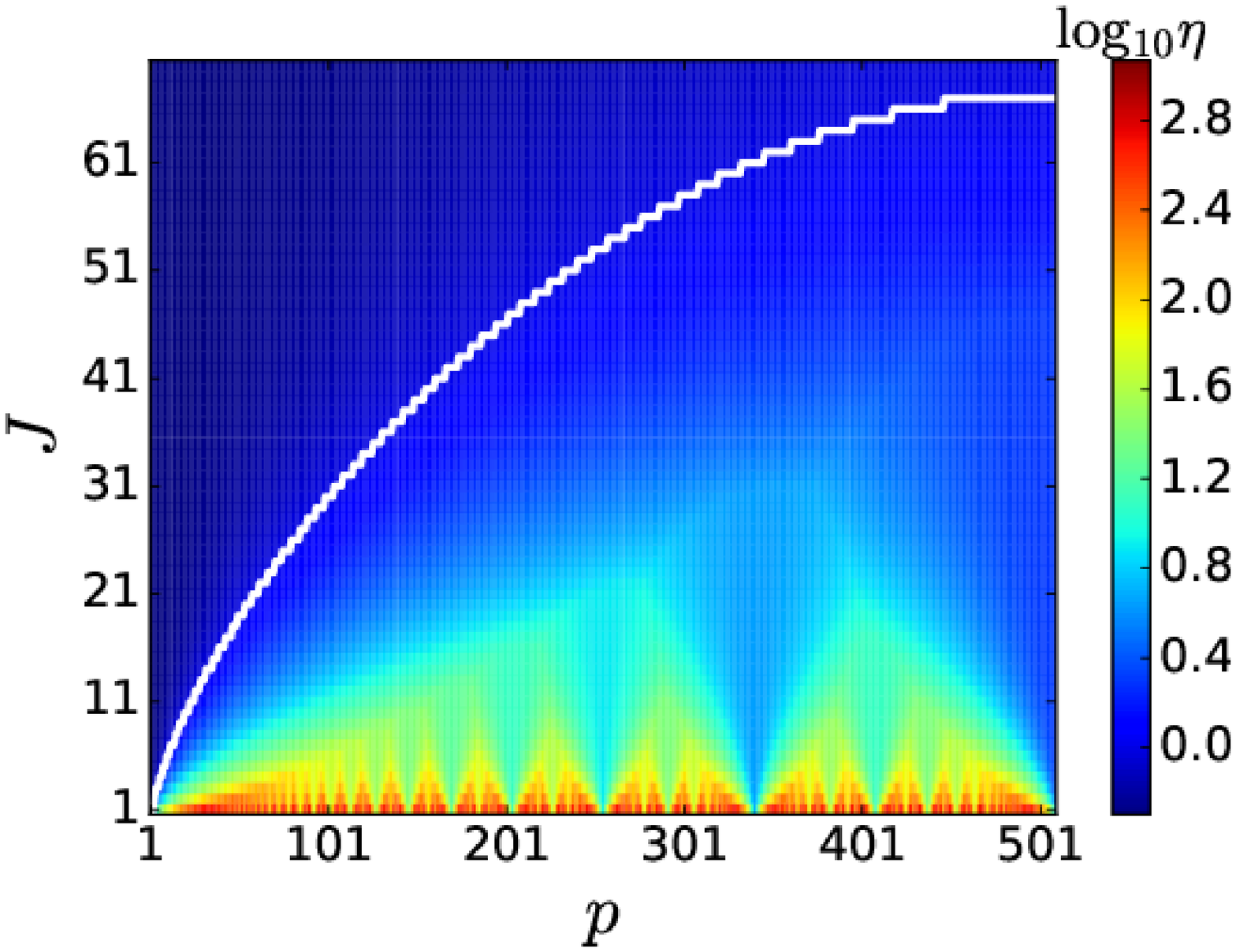}
\label{}
}
\subfloat[]{
\centering
\includegraphics[width=0.5\linewidth]{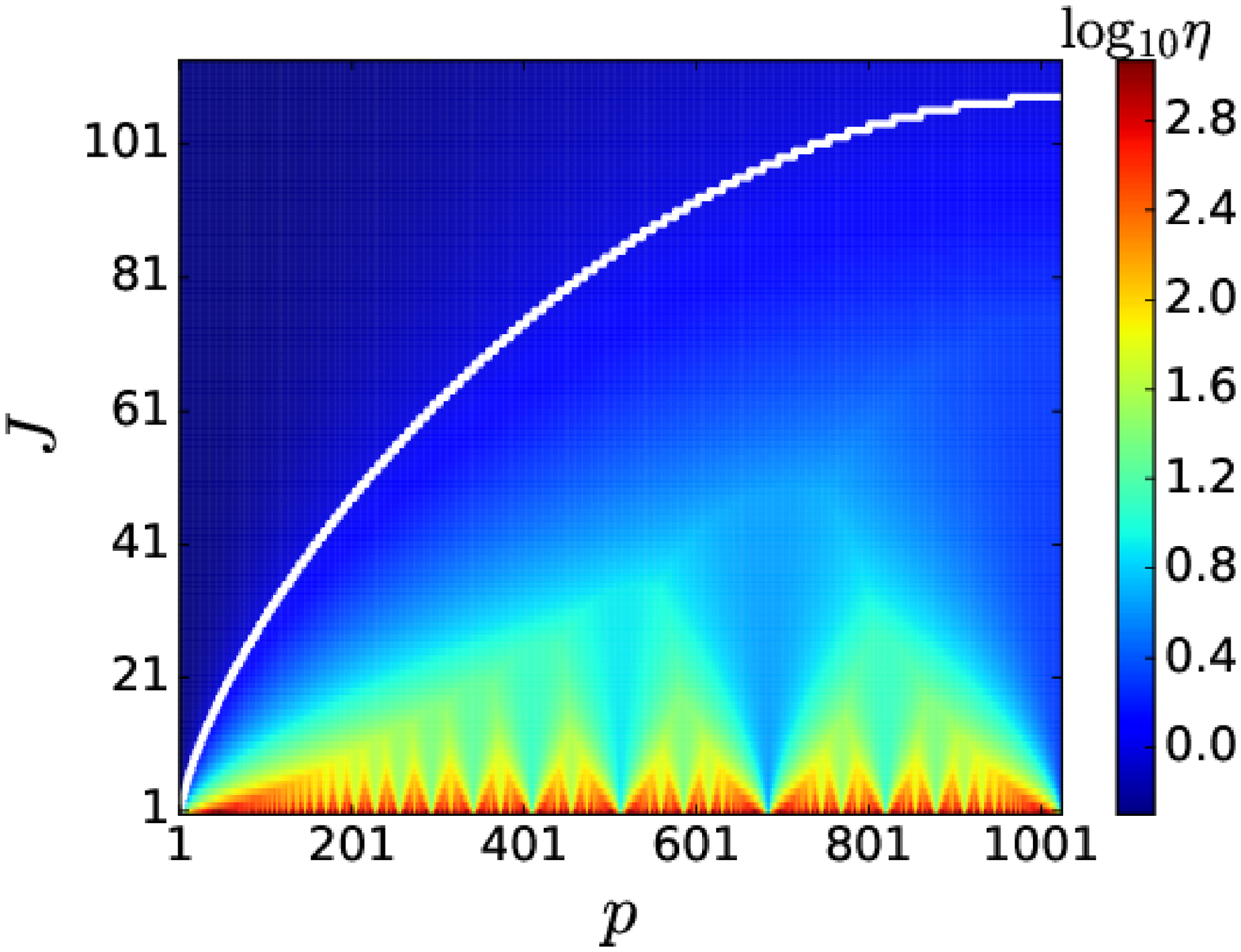}
\label{}
}
\caption{Dependence of $\eta$ on $J$ and $p$ for multiplex prime stars with $N=1021$ (a) and $N=2053$ nodes (b), respectively,when $D_x\rightarrow \infty$. White line represents the limit between superdiffusive and normal diffusive behaviors ($\eta=1$).}
\label{super_diff}
\end{figure}

In Fig.~\ref{super_diff} we show an example of the dependence of $\eta$ on $J$ and $p$, when $D_x\rightarrow \infty$ (i.e., $\Lambda_2\rightarrow\Lambda_2^A$). As can be observed, for a given system size $N$, there is a threshold value of $J$, denoted as $J_x$, such that, if $J>J_x$, super-diffusion is hindered, regardless of the value of $p$. When $p=(N-1)/2$, we found numerically that the natural number that minimizes (\ref{asin_eigenB}) is $c=2$. On the assumption that $\eta=1$, $J_x$ meets the following condition:

\begin{align}
\left (J_x-1-\cos\left (\frac{\pi}{N}\right )\right )\sin\left (\frac{\pi}{N}\right )=\sin\left (\left (2J_x+1\right )\frac{\pi}{N}\right )-\sin\left (\left (J_x+1\right )\frac{\pi}{N}\right )\cos\left (J_x\frac{\pi}{N}\right )
\label{def_J_x}
\end{align}

\noindent For $N\gg1$, the result of (\ref{def_J_x}) can be approximated by $J_x\approx 0.672 N^{2/3}$. According to (\ref{def_J_x}), $J_x$ only depends on $N$, and, thus, the larger the value of $N$, the larger $J_x(N)$. For instance, $J_x(N=1021)\approx67.88$ for and $J_x(N=2053)\approx108.31$ (see Fig.~\ref{super_diff}). Therefore, unlike the results described for regular random graphs in \cite{cencetti19}, here super-diffusion does not necessarily emerge when the edge overlap meets the condition $w<1$ [see (\ref{edge_w_interac})], even in the limit case of $D_x\rightarrow \infty$.

On the other hand, in Fig.~\ref{super_diff} it can also be seen that the larger values of $\eta$ appear when $J=1$. As expected, the addition of extra-links to the interacting cycle speeds up the diffusion in that layer and, consequently, it ruins the beneficial effect that the multiplexity may have on diffusion time scale. Besides that, we can observe that superdiffusive values of $\eta$ recursively reach a minimal values when $J\sim1$. The first one occurs at the largest possible value of $p=(N-1)/2$, which is clearly followed by a second minimum at $p\approx(N-1)/3$, and the indication of a third event at $p\approx(N-1)/4$. We remark that the event at $p\approx(N-1)/3$ coincides with the minimum value of $\delta$ indicated by arrows in both panels of Fig.~\ref{Lambda_2_fig}. Finally, as observed previously in Fig.~\ref{delta_Jp}, a self-similar pattern is exhibited in both panels of Fig.~\ref{super_diff} after a linear scaling of the $(J,p)$ combinations with respect to $N$. Indeed, according to the approximate solution of (\ref{def_J_x}), the linear scaling of $J$ is simply given by $J_1\approx J_2 J_x(N_1)/J_x(N_2)\approx J_2 (N_1/N_2)^{2/3}$, when considering two prime stars multiplex with sizes $N_1$ and $N_2$, respectively, and $D_x\rightarrow \infty$.

\section{Conclusions}
\label{Sec:conclusions}

In this work we investigated a simple, mathematically tractable set of multiplexes based on individual layers with quite similar structures. Despite such similarity and provided the number of vertices is a prime number, the nodes in each layer can be labeled in different ways. Each labeling choice leads to multiplex with different properties, which are reflected by a reliable measure for layer dissimilarity and by the properties of diffusive processes on the multiplex. The fact that the relevant matrices for the multiplex structure and dynamical processes still hold the circulant property is the key element for the derivation of analytical results presented here. The numerical evaluation of the resulting expressions indicate consistent scaling properties as a function of the number of nodes in the network and the pair $(J,p)$ of parameters defining the multiplex. Although we restricted the presentation of results for two-layer multiplexes, their general mathematical grounds are valid for larger values of $M$.


\begin{acknowledgments}

This work was supported by the Brazilian agencies CAPES and CNPq through the Grants 151466/2018-1 (AA-P) and 305060/2015-5 (RFSA). RFSA also acknowledges the support of the National Institute of Science and Technology for Complex Systems (INCT-SC Brazil).

\end{acknowledgments}




\begin{thebibliography}{99}
\bibitem{note1}
Since interacting cycle graphs and prime stars are connected graphs and the multiplex has only one connected component, the 0 eigenvalue of the supra-Laplacian matrix only appears once (i.e., 0 has algebraic multiplicity 1).
\end{thebibliography}
\end{document}